\documentclass[aps,letterpaper,superscriptaddress]{revtex4}
\usepackage{epsfig}
\def\be{\begin{equation}}
\def\ee{\end{equation}}
\def\bea{\begin{eqnarray}}
\def\eea{\end{eqnarray}}

\def\nn{\nonumber}

\def\rt{\rightarrow}
\newcommand{\DIR}{./}
\begin{document}
\title{\bf Motion of a random walker in a quenched power law correlated 
velocity field}

\author{Soumen Roy}
\affiliation{Institute of Physics, Bhubaneswar 751005, India}
\email{sroy@iopb.res.in}
\author{Dibyendu Das}
\affiliation{Department of Physics, Indian Institute of Technology Bombay, 
Powai, Mumbai 400076, India}
\email{dibyendu@phy.iitb.ac.in}

\begin{abstract}

We study the motion of a random walker in one longitudinal and $d$
transverse dimensions with a quenched power law correlated velocity
field in the longitudinal $x$-direction. The model is a modification
of the Matheron-de Marsily (MdM) model, with long-range velocity
correlation. For a velocity correlation function, dependent on
transverse co-ordinates $\vec y$ as $1/(a+|{\vec{y}_1 -
\vec{y}_2}|)^{\alpha}$, we analytically calculate the two-time
correlation function of the $x$-coordinate.  We find that the motion
of the $x$-coordinate is a fractional Brownian motion (fBm), with a
Hurst exponent $H = {\rm max}[1/2, (1-\alpha/4), (1-d/4)]$.  From this
and known properties of fBM, we calculate the disorder averaged
persistence probability of $x(t)$ up to time $t$.  We also find the
lines in the parameter space of $d$ and $\alpha$ along which there is  
marginal behaviour.  We present results of simulations which support our
analytical calculation.
\noindent

\end{abstract}

\maketitle

\section{Introduction}

It is well known that a random walker making short-range hops, namely
a Brownian particle in a thermal environment, has an average
incremental displacement which goes as square root of the time
difference. Although this property is rather robust, deviation from
this diffusive behaviour has been observed in a wide variety of
physical situations \cite{BG}.  Super-diffusion is well known in the
context of turbulence and L$\acute{\rm e}$vy walks \cite{turbulence}.  Yet
another way to see anomalous diffusion is to subject the random walker
to a quenched velocity field. A concrete example of this is the
Matheron-de Marsily (MdM) model, which was introduced to study the
hydrodynamic dispersion of solute particles in sedimentary layered
porous rock formation \cite{MM}. In the MdM model a single particle
diffuses in a $(1+1)$-dimensional layered medium, such that the
disorder is anisotropic. The motion of the particle is purely Brownian
along the transverse $y$ direction, while along the longitudinal $x$
direction, added to the thermal noise, the particle is driven by a
quenched drift velocity $v(y)$ that is a random function of only the
transverse coordinate $y$. The coupling between diffusion and
convection due to the spatially random but temporally static velocity
field, generates a typical bias in the $x$ direction giving rise to a
super-diffusive longitudinal transport for large $t$
\cite{BG,BJKP,ZKB,R1}. 

The temporal behaviour is quantitatively described by the standard
two-time dependent correlation function, and the persistence
probability. Intuitively, persistence of a system concerns the
property of the system to stay in some given state.  The persistence
probability $P(t)$ which is related to the first-passage probability
$F(t)$, as $F(t)=-dP(t)/dt$ is simply the probability that the
particle does not cross a given point up to time $t$. In systems with
super-diffusion, $P(t)$ is expected to get enhanced. Persistence
properties in disordered systems are harder to find, and thus results
known for such models like the Sinai model \cite{Sinai,Sinai1,Sinai2}
and the MdM model \cite{R1,R2} are of great importance. Recently for a
$(d+1)$ dimensional version of MdM model the persistence probability
was analytically derived in \cite{sn1}.

In earlier studies of the MdM model \cite{BJKP,R1,R2,sn1}, only short
range correlated $v(y)$ had been considered. However, since originally
the model was motivated \cite{MM} by general flow in certain types of
fractured rocks with parallel fractures which allow the propagation of
dissolved species, it is possible that the flows parallel to the
bedding might have velocity correlations which are spread over long
distances. It may be mentioned that different aspects of dynamics in 
MdM fields with algebraic correlations, have been studied  \cite
{dynamics}.  Generally long range correlated disordered problems are
hard to solve. For isotropic disorder with power law correlations,
various results are known \cite{BG,konk,B3}. The MdM model being
anisotropic allows for exact anatytic treatment as compared to field
theoretic treatment done in other cases.  In this paper we have been
able to analytically derive the incremental correlation function and
persistence probability, for the MdM model with long range {\it power
law} correlated $v(y)$. We find that long range correlation
incorporates new temporal behaviour, as compared to the short range
MdM model. The relevant exponent of $P(t)$ becomes dependent on the
power law exponent $\alpha$ and the transverse dimension $d$.

The method that we follow is similar to \cite{sn1,sn2} and was used 
even earlier in \cite{Krug1}. We first show 
that $x(t)$ is a fractional Brownian motion (fBm), and then use its
known first passage property to solve for the persistence probability.
A stochastic process $x(t)$ (with zero mean $\langle x(t) \rangle =
0$) is called a fBm if its incremental two-time correlation function
$C(t_1,t_2) = \langle{\left(x(t_1)-x(t_2)\right)^2}\rangle$ is 
stationary, i.e., depends only on the difference $|t_1-t_2|$ and
moreover, grows asymptotically as a power law,
\begin{equation}
C(t_1,t_2) \sim |t_1-t_2|^{2H}, \,\,\,\, |t_1-t_2|>>1. 
\label{hurst1}
\end{equation}
The parameter $H$ is called the Hurst exponent that characterizes the
fBm\cite{MvN} and $\langle \dots \rangle$ denotes the expectation
value over all realizations of the process. For example, the ordinary
Brownian motion is a fBm with $H=1/2$. The zero crossing properties of
fBm has been studied before\cite{Krug1,B,HEM,DW} and it is known, both
analytically and numerically, that the probability that a fBm does not
cross zero up to time $t$ has a power law decay, $P(t)\sim
t^{-\theta}$ for large $t$ with $\theta=1-H$\cite{Krug1,DW}.  We use
this property to derive $\theta$ from $H$.

\section{Model and main results}

We consider a generalized version of the MdM model in $d (> 0)$
transverse plus $1$ longitudinal spatial dimensions. The transverse
coordinates of the particle, denoted by $y_i(t)$'s with $i=1,2,\dots,
d$, undergo ordinary Brownian motion, 
\begin{equation}
{\dot y_i}=\eta_i(t),
\label{yc}
\end{equation}
where $\eta_i$'s are zero mean Gaussian white noises with 
correlators, $\langle \eta_i(t)\eta_j(t')\rangle=\delta_{i,j}\delta (t-t')$. 
Equivalently we can use a single equation involving $d$-dimensional vectors
${\vec y} =\{y_i(t)\}$ and ${\vec \eta} =\{\eta_i(t)\} $:
\begin{equation}
{\dot {\vec y}}={\vec \eta}(t).
\label{yc1}
\end{equation}
The longitudinal coordinate $x(t)$, in contrast, 
is driven by a random drift $v[{\vec y}(t)]$ that depends only on the 
transverse coordinates ${\vec y}(t)$,
\begin{equation}
{\dot x}= v[{\vec y} (t)] +\eta_x(t),
\label{xc}
\end{equation}
where $\eta_x$ is again a zero mean Gaussian white
noise with $\langle \eta_x(t_1)\eta_x(t_2)\rangle=\delta(t_1-t_2)$.  
The velocity field $v[\vec y]$ is quenched, i.e., constant in time.  
Operationally this means that for a given realization of the function $v$, 
one has to evolve $x(t)$ over `thermal histories' in Eq. (\ref{xc}) and
then one needs to `disorder average' (denoted by ${\overline
{\dots}}$) over different realizations of the random function $v$. In
the original MdM model and in the latter works on slightly modified
versions of it \cite{BG,MM,BJKP,ZKB,R1,R2,sn1}, 
the random function
$v(\vec{y})$ has been considered to be a Gaussian with zero mean 
(${\overline {v(\vec y)}}=0$), with a correlator ${\overline {v(\vec y_1)
v(\vec y_2)}}$, 
which is either $\delta(\vec y_1-\vec y_2)$ or is short ranged namely
${1/{[2\pi d a^2]^{d/2}}} \exp\left[-{ {\left(\vec y_1-\vec
y_2\right)^2}/{2 d a^2}}\right]$ (where $a$ is a small cutoff).

In contrast to the above, here we consider the random function $v$ 
to have long-range power-law correlation:
\bea
{\overline {v(\vec y)}} &=& 0 \nn \\
{\overline {v(\vec y_1) v(\vec y_2)}} &=& {1 \over {(a + |\vec y_1 - 
\vec y_2|)^{\alpha}}}.
\label{vv} 
\eea 
The exponent $\alpha > 0$ denotes the strength of the correlation
function, and the short distance cutoff scale `$a$' ensures that there
is no divergence as $|\vec y_1 - \vec y_2| \rt 0$. This model is
analytically treated in Section \ref{ana} and numerically in Section
\ref{num}. We will see that the role played by the scale $a$ is very
interesting. For $\alpha < d$, it is irrelevant and can be set to
zero, but for $\alpha > d$, it has to be finite to ensure a
well-behaved asymptotic behaviour.

Our main result is that the thermal and disorder averaged two-time
correlation function of the $x$-coordinate,  $ \overline{\langle
\left(x(t_1)-x(t_2)\right)^2 \rangle} \approx |t_2 - t_1|^{2H}$. Thus
the process $x(t)$ is an fBM. The Hurst exponent $H = {\rm max}[1/2,
(1-\alpha/4), (1-d/4)]$. This means that for $d < 2$, the exponent
will change from $H = (1-\alpha/4)$ for $\alpha < d$, to $H = (1-d/4)$
for $\alpha > d$. On the other hand, for $d \geq 2$, $H =
(1-\alpha/4)$ for $\alpha < 2$, and $H = 1/2$ for $\alpha > 2$.  It
follows from this that the persistence probability of $P(t) \sim
t^{-\theta}$, with $\theta = 1 - H = {\rm min}[1/2, \alpha/4, d/4]$.
We have also found the lines in $d$ and $\alpha$ plane along which
there is marginal behaviour.

\section{Two-Time Correlation function and Persistence probability}
\label{ana}

The analytical treatment that we present here is similar to that of
\cite{sn1,sn2} up to the point that the explicit power law form of
velocity correlation is substituted. Nevertheless, for completeness,
we give the steps briefly.  In order to solve for correlation
functions of $x(t)$ from Eq. \ref{xc}, we need knowledge of
correlation functions of $\vec y(t)$.  By integrating Eq. \ref{yc1} we
get
\begin{equation}
{\vec y}=\int_{0}^{\tau}{\vec \eta}(\tau)d\tau
\label{y1}
\end{equation}
From the above linear relation we conclude that since $\vec \eta$ is a
Gaussian random variable, so is $\vec y$. Moreover another vector
${\vec Y}={\vec y}(\tau_1)-{\vec y}(\tau_2)$ which will be useful in
the following calculations is also a Gaussian random variable by virtue 
of the linear relation. Using Eq. \ref{y1},
\begin{equation}
{\vec Y}=\int_{0}^{\tau_1}{\vec \eta}(\tau)d\tau-\int_{0}^{\tau_2}{\vec \eta}(\tau)d\tau,
\label{Y1}
\end{equation}
and it follows that $\langle {\vec Y}\rangle=0$ and variance 
$\langle {\vec Y}^2 \rangle= d|\tau_1-\tau_2|$, since 
$\langle \eta_i(t)\eta_j(t')\rangle=\delta_{i,j}\delta (t-t')$. 
Thus explicitly the distribution 
\begin{equation}
P(\vec Y)=[2\pi d|\tau_1-\tau_2|]^{-d/2}\exp\left[{-{\vec Y}^2/{2d|\tau_1-\tau_2|}}\right].
\label{Gauss}
\end{equation}

With the above solution of the transverse coordinates $\vec y(t)$, we 
now proceed to treat the longitudinal coordinate $x(t)$. We note that the 
averages so far over the transverse coordinates, namely $\langle \dots 
\rangle$ were thermal averages. The coordinate $x(t)$ depends on both a 
thermal noise $\eta_x$ and a quenched random noise $v$. So we need to 
do a thermal as well as quenched disorder average, which we will denote 
by $\overline {\langle \dots \rangle}$, where in particular the $\overline{ 
\dots}$ denotes the quenched disorder average.    
Integrating Eq. (\ref{xc}) we get, 
\begin{equation}
x(t)= \int_0^t \eta_x(\tau)d \tau + \int_0^t v[\vec y(\tau)]d\tau.
\label{xc1}
\end{equation}
From Eq. (\ref{xc1}) we get $\overline {\langle x(t)\rangle}=0$, while using 
$\langle \eta_x(t_1)\eta_x(t_2)\rangle=\delta(t_1-t_2)$, we get 
\begin{equation}
\overline{\langle x(t_1)x(t_2)\rangle}= {\rm {min}}(t_1,t_2)+I(t_1,t_2),
\label{xsol1}
\end{equation} 
where the integral $I(t_1,t_2)$ is given by
\begin{eqnarray}
I(t_1,t_2)&=& \int_0^{t_1}\int_0^{t_2} \langle 
{\overline { v[{\vec y}(\tau_2)]v[{\vec y}(\tau_2)]} }\rangle d\tau_1d\tau_2 \nonumber \\
&=&\int_0^{t_1}\int_0^{t_2} \left\langle {1 \over {(a + |\vec Y|)^{\alpha}}} \right\rangle d\tau_1 d\tau_2.
\label{Idef}
\end{eqnarray}
In the above, we have used the correlator of $v$ from Eq. \ref{vv} and 
$\vec Y$ is the same as in Eq. \ref{Y1}. Since we know the probability 
distribution $P(\vec{Y})$ explicitly from Eq. \ref{Gauss}, we can  
proceed from Eq. \ref{Idef} to evaluate
\bea
I(t_1,t_2) &=& \int_{0}^{t_1} \int_{0}^{t_2} \int_{-\infty}^{\infty} d\tau_1 d\tau_2 {1 \over (|\vec Y| + a)^\alpha} {e^{-{|\vec Y|^2 \over {2d|\tau_1 - \tau_2|}}} \over {(2\pi d |\tau_1 -\tau_2|)^{d/2}}} d^d{\vec Y} \nn \\
&=& {2 \over \Gamma(d/2)} \int_{0}^{t_1} \int_{0}^{t_2} \int_{0}^{\infty} 
{{d\tau_1 d\tau_2 e^{-Z^2} Z^{d-1} dZ} \over {(Z \sqrt{2d |\tau_1 - \tau_2|} + 
a)^{\alpha}}}  \nn \\
  &=& {2 \over \Gamma(d/2)} \int_{0}^{\infty}  e^{-Z^2} Z^{d-1} ~ J(Z,t_1,t_2) ~ dZ.
\label{It1t2}
\eea
In Eq. \ref{It1t2} we have substituted $|\vec Y| = Z 
\sqrt{2d |\tau_1 - \tau_2|}$ in going from the first line to the second, 
and then performed the two time integrals exactly to get explicitly 
the function  
\bea
J(Z,t_1,t_2) &=& \int_{0}^{t_1} \int_{0}^{t_2} {{d\tau_1 d\tau_2} \over 
{(Z \sqrt{2d |\tau_1 - \tau_2|} + a)^\alpha}} \\
\label{Jdef}
&=& {1\over {Z^2 d}}{{2 a^{2-\alpha} {\rm min}(t_1,t_2)} \over {(2-\alpha)(1-\alpha)}} + {{(a+Z\sqrt{2 t_1 d})^{4-\alpha} + (a+Z\sqrt{2 t_2 d})^{4-\alpha} - a^{4-\alpha} - (a+Z\sqrt{2d |t_2 - t_1|})^{4-\alpha}} \over {Z^4 d^2 (2-\alpha) (4 - \alpha)}} \nn \\
&+& {{(3-2\alpha) \left[ 
- a(a+Z\sqrt{2 t_1 d})^{3-\alpha} -  a(a+Z\sqrt{2 t_2 d})^{3-\alpha} + a^{4 - \alpha} + a(a+Z\sqrt{2 d |t_2 - t_1|})^{3-\alpha} \right]} \over {Z^4 d^2 (3-\alpha) (1 - \alpha) (2 - \alpha)}}  \nn \\
&+& {{a^2(a+Z\sqrt{2 t_1 d})^{2-\alpha} + a^2(a+Z\sqrt{2 t_2 d})^{2-\alpha} - a^{4-\alpha} - a^2(a+Z\sqrt{2d |t_2 - t_1|})^{2-\alpha}} \over {Z^4 d^2 (1-\alpha) (2 - \alpha)}}.
\label{Jfull}
\eea

Although Eq. \ref{Jfull} has many terms, the two time-correlation
function $C(t_1,t_2) = \overline{\langle x^2(t_1)\rangle} +
\overline{\langle x^2(t_2)\rangle} - 2 \overline{\langle
x(t_1)x(t_2)\rangle}$ has a dependence only on $t = |t_1 - t_2|$, and has
the following comparatively simpler expression: 
\bea C(t) = t ~ + ~ 
{4 \over \Gamma(d/2)} \int_{0}^{\infty} dZ~ e^{-Z^2} Z^{d-1} \{ {{A t} \over
Z^2} + {1 \over Z^4} &[& 2B(a+Z\sqrt{2d t})^{4-\alpha} - 2 C
a(a+Z\sqrt{2d t})^{3-\alpha} \nn \\ &+& 2D a^2 (a+Z\sqrt{2d
t})^{2-\alpha} - (2B-2C+2D)a^{4-\alpha}]\}
\label{C}  
\eea 
In Eq. \ref{C} we have used the symbols $A = 2
a^{2-\alpha}/[d(2-\alpha) (1-\alpha)]$, $B =
1/[d^2(2-\alpha)(4-\alpha)]$, $C = (3-2\alpha)/[d^2
(1-\alpha)(2-\alpha)(3-\alpha)]$ and $D = 1/[d^2(1-\alpha)(2-\alpha)]$. 
Further, Eq. \ref{C} can be written as $C(t) = t + \bar{C}$, where 
$\bar{C}$ is the integral which has to be analyzed to get the leading 
temporal behaviour.   

Defining a new variable of integration $y^{\prime} = Z \sqrt{2 d t}$ and substituting in $\bar{C}$, we get 
\bea
\bar{C} = {{4~ t^{2-{d/2}}} \over {\Gamma(d/2) (2d)^{d/2}}} \int_{0}^{\infty} d{y^{\prime}}~ e^{-{{y^{\prime}}^2/{2d t}}}~ {y^{\prime}}^{d-1} \{ {{A 2d} \over {y^{\prime}}^2} + {{4d^2} \over {y^{\prime}}^4} &[& 2B(a+y^{\prime})^{4-\alpha} - 2 C a(a+y^{\prime})^{3-\alpha} \nn \\
&+& 2D a^2 (a+y^{\prime})^{2-\alpha} - (2B-2C+2D)a^{4-\alpha}]\}
\label{Cbar}
\eea 

It turns out that one has to be very careful to extract the asymptotic
behaviour of the above integral, because the correlations at scales
$y^{\prime} < a$ and $y^{\prime} > a$ contribute different power laws,
and they compete with each other. It is best to split the integral
$\int_{0}^{\infty}$ in Eq. \ref{Cbar} into two parts $\int_{0}^{a}$
and $\int_{a}^{\infty}$. Accordingly $\bar{C} = C_1 + C_2$. In $C_1$
we expand the terms of the integrand in powers of $y^{\prime}/a$ (with
$y^{\prime} < a$), while in $C_2$ we expand the terms in powers of
$a/y^{\prime}$ (with $y^{\prime} > a$).

The $C_1$ integral, with its integrand expanded in a series yields: 
\bea
C_1 &=& {{4~ t^{2-{d/2}}} \over {\Gamma(d/2) (2d)^{d/2}}} \int_{0}^{a}
dy^{\prime}~ e^{-{{y^{\prime}}^2/{2d t}}}~ {y^{\prime}}^{d-1} ~ a^{-\alpha} ~[1 - {4\alpha \over 15}
{{y^{\prime}} \over a} + O({{y^{\prime}}^2 \over a^2})] \nn \\
&\approx& b_1 a^{d-\alpha} ~ t^{2-{d/2}} 
\label{C1}
\eea 
where $b_1 = {{4} \over {\Gamma(d/2) (2d)^{d/2}}} [ {1 \over d}  
- {{4 ~\alpha } \over {15 (d+1)}} + \cdots ]$.
In obtaining Eq. \ref{C1} we have used the fact that for small $y^{\prime}$ and
very large $t$, we may approximate $e^{-{y^{\prime}}^2/{2dt}} \approx 1$ inside 
the integral. In Eq. \ref{C1} the factor $a^{d-\alpha}$ smoothly 
goes to zero if $a \rt 0$ for $\alpha < d$. If we wish, by taking smaller
and smaller cutoff $a$, we can actually ignore the contribution of 
of $C_1$ with respect to $C_2$. On the other hand, if $\alpha > d$, 
$a \rt 0$ will lead to a divergence, and hence we have to keep $a$ finite, 
and thus the contribution of $C_1$ cannot be ignored. 
 
\begin{figure}
\begin{center}
\psfig{file=\DIR/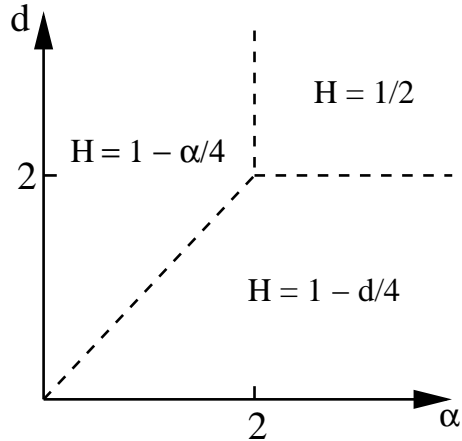,width=6.0cm,angle=0}
\caption{The values of the exponent $H$ in the three regions of the 
parameter space of $d$ versus $\alpha$ are shown.}
\label{fig4}
\end{center}
\end{figure}

Now we turn to the integral $C_2$ and it goes as follows:
\bea
C_2 &=& {{4 t^{2-{d/2}}} \over {\Gamma(d/2) (2d)^{d/2}}}
\int_{a}^{\infty} dy^{\prime}~ e^{-{{y^{\prime}}^2/{2d t}}}~ {y^{\prime}}^{d-1} \{{{A 2d} \over {y^{\prime}}^2} + {{4d^2} \over {y^{\prime}}^{\alpha}} [ 2B + (2B (4 - \alpha) - 2C)~{a \over {y^{\prime}}} + O({a^2 \over {y^{\prime}}^2}) - (2B - 2C + 2D)~{a^{4-\alpha} \over {y^{\prime}}^{4-\alpha}} ]\} \nn \\ 
&\approx&  a_0 ~ t + a_1 ~ t^{2-\alpha/2} + O(t^{3/2-\alpha/2}) 
+ b_2 a^{d - \alpha} ~ t^{2-{d/2}}
\label{C2} 
\eea 
where the newly defined constants are $a_0 = {{4 A}/
[{\Gamma(d/2)(d-2)}}]$, $a_1 = {{32 B d^2}/[{\Gamma(d/2)(d-\alpha)
(2d)^{\alpha/2}}}]$, $b_2 = {4 \over
\Gamma(d/2)}\{-{{4}/[{(d-2)(1-\alpha)(2-\alpha)}}] -
{{8Bd^2 }/(d-\alpha)} + O(a) + {b_4}\}$, and $b_4 =
24/[(d-4)(1-\alpha)(2-\alpha)(3-\alpha)(4-\alpha)]$. In going from the
first line to the second line in Eq. \ref{C2} we have made the
approximation of replacing the upper limit $\infty$ by $\sqrt{2d t}$,
as the exponential factor $exp(-{{y^{\prime}}^2/{2d t}})$ sharply cuts off the
integral beyond that scale anyway. After that we have set
$e^{-{y^{\prime}}^2/{2dt}} \approx 1$ and then done the integrals. We see that in
Eq. \ref{C2}, $C_2$ has three dominant power laws with powers equal to
$1$, $(1 - \alpha/2)$, and $(1 - d/2)$. Combining Eqs. \ref{C1} and
\ref{C2}, we get finally the two-time incremental correlation function
from Eq. \ref{C} as 
\bea 
C(t_1,t_2) = C(t) \approx B_1 t + a_1 t^{2 - \alpha/2} + B_2 t^{2
- d/2},
\label{Cfinal}
\eea 
where $B_1 = 1 + a_0$, and $B_2 = (b_1 + b_2) a^{d-\alpha}$.  This
implies that asymptotically, at long time $t$, $C(t) \sim t^{2H}$.
Thus $x(t)$ is a fBM process and the  associated 
Hurst exponent $H$ is given by 
\bea 
H = {\rm max}[1/2, (1 - \alpha/4), (1 - d/4)]
\label{hurst2} 
\eea 
More explicitly, it means that for $d < 2$, for $\alpha < d$, $H
= (1 - \alpha/4)$ while for $\alpha > d$, $H = (1 - d/4)$. On the
other hand, for $d \geq 2$, for $\alpha < 2$, $H =(1 - \alpha/4)$
while for $\alpha > 2$, $H = 1/2$. This scenario is shown in Fig. 
\ref{fig4}.

We have analysed the cases of marginal behaviour separately.  For
short-range disordered MdM model, $d=2$ was shown \cite{sn1} to have
marginal behaviour. For the current model, we have done explicit
calculations (similar as above) for the cases $\alpha = 2$, $d=2$ and
$\alpha = d$ \cite{ddas}.  $(i)$ For $\alpha = 2$, for large $t$,
$C(t) \sim t ~{\rm ln}~ t$ (for $d > 2$), ~$\sim t^{2 - d/2}$ (for $d < 2$), 
and $\sim t~ ({\rm ln}~t)^2$ (for $d=2$). This means that
the semi-infinite line $\alpha = 2$ with $d \geq 2$ (see
Fig. \ref{fig4}) has marginal behaviour. $(ii)$ For $d = 2$, $C(t)
\sim t~ {\rm ln}~t$ (for $\alpha > 2$) and $\sim t^{2 - \alpha/2}$ 
(for $\alpha < 2$). Hence the semi-infinite line $d=2$ with
$\alpha > 2$ (Fig. \ref{fig4}) exhibits marginality.  $(iii)$ Finally
we also find logarithmic correction, namely $C(t) \sim t^{2 - d/2}
{\rm ln}~t$, along the line $\alpha = d$, for $\alpha$ and $d$ both $< 2$.

From the known first-passage property of fBm mentioned earlier, the
disorder and thermal averaged probability that the process $x(t)$ does
not cross zero up to time $t$ decays as a power law, $P(t) \sim
t^{-\theta}$ with $\theta=1-H$. Using the results for $H$ in
Eq. (\ref{hurst2}), we get 
\bea
\theta = {\rm min}[1/2, \alpha/4, d/4]. 
\label{theta}
\eea
This means that for  $d < 2$, if $\alpha < d$, $\theta = \alpha/4$ 
while if $\alpha > d$, $\theta = d/4$. For $d \geq 2$, if $\alpha < 2$, 
$\theta = \alpha/4$, while if $\alpha > 2$, $\theta =  1/2$. 

Although for $d < 2$ and $\alpha > d$, we expect analytically $C(t)
\sim t^{2-d/2}$ for large $t$, we find numerically (see section
\ref{num}) that it is rather tricky to find this power law.  The
latter is a consequence of the Eq. \ref{C1}, which tells us that this
long time behaviour arises, predominantly, as a contribution of
correlations at distance scales $y^{\prime} < a$. On the other hand
Eq. \ref{C2} shows that the power law $t^{2 - \alpha/2}$ arises out of
a contribution due to correlations at scale $y^{\prime} > a$. So if a
finite cutoff $a$ is not kept, and the distance scales $y^{\prime} \ll a$
are not probed carefully, numerically one may find a wrong asymptotic
power namely $2H = (2 - \alpha/2)$ instead of the expected power $2H =
(2 - d/2)$. We will discuss this further in the next section
\ref{num}.

\section{Numerical simulation and results}
\label{num}

Since the exponents in Eq. \ref{hurst2} followed from Eq. \ref{Cfinal}
which is an approximate asymptotic form of the exact Eq. \ref{C}, we
first checked by numerically integrating Eq. \ref{C} that the
predicted powers are indeed true. While doing this, we realized that
in order to see the power $C(t) \sim t^{2-d/2}$ in the regime $\alpha
> d$ (for $d<2$) one should be very careful to choose the numerically 
finite integration measure $\delta Z \ll 1$ (where $\delta Z$  
approximates infinitesimal $dZ$), in order to see the asymptotics.
This is related to the comment made earlier that we have to probe
$y^{\prime} \ll a$ (where $y^{\prime} = Z \sqrt{2 d t}$ as mentioned 
earlier) to get the asymptotic contribution, in this regime.

We now turn to verifying the analytical predictions of the 
previous section, namely $H$ as in Eq. \ref{hurst2} and the corresponding 
exponent $\theta$ as in Eq. \ref{theta}, by doing numerical simulations. 
We use the time discretised version of Eqs. \ref{xc} and \ref{yc}, 
\begin{eqnarray}
y_i(t_{m+1})&=& y_i(t_m) + \sqrt{\Delta t} \eta_i(t_m) \label{dis1} \\
x(t_{m+1})&=&x(t_m)+ {\Delta t}\, v\left(x(t_m)\right) + 
\sqrt{\Delta t}\,\eta_x(t_m), \label{dis2}
\end{eqnarray}
where $t_m=m {\Delta t}$, with $m$ being an integer.    
We choose $\Delta t < 0.5$  in our simulations so that the 
stability is guaranteed \cite{Krug1}. The variables $\eta_i(t_m)$ and 
$\eta_x(m)$ are independent Gaussian variables for all $t_m$ and each is 
distributed with zero mean and unit variance. Moreover our simulations here 
are all for $d=1$, i.e. we have a single transverse coordinate $y$. Hence 
the subscript $i$ assumes a single value equal to $1$.  

Unlike \cite{sn1,sn2}, here we had to numerically create quenched power law 
correlated velocity $v(y)$ of the form given by Eq. \ref{vv}. 
We note that if we have a delta-correlated 
velocity correlation in Fourier $k$-space with a $k$-dependent prefactor  
as follows
\be
\overline{{\tilde v}(k_1) {\tilde v}(k_2)} = c(ka) 
|k_1|^{\alpha -1} \delta(k_1 + k_2),     
\label{vv_k}
\ee 
with constant $c(ka) = [{\rm cos}(ka) \int_{ka}^{\infty} dp {{\rm
cos}p \over p^{\alpha}} + {\rm sin}(ka) \int_{ka}^{\infty} dp {{\rm
sin}p \over p^{\alpha}}]$ , in the real $y$-space we have $\overline{v(y_1)
v(y_2)}$ given by Eq. \ref{vv} up to an constant factor. The factor
$c(ka) |k_1|^{\alpha -1}$ in Eq. \ref{vv_k} is the square of the
variance of the distribution of $\tilde v$ in $k$-space.  In our
numerics, we first generate a realization of Gaussian distributed
random numbers ${\tilde v}(k)$ in $k$-space with zero mean and
variance proportional to $\sqrt{c(ka)} |k|^{(\alpha - 1)/2}$, and then
Fourier transform back to real space.  The latter realization of
random $v(y)$ are thus guaranteed to have the required power law
correlation in real $y$-space.

\begin{figure}
\begin{center}
\psfig{file=\DIR/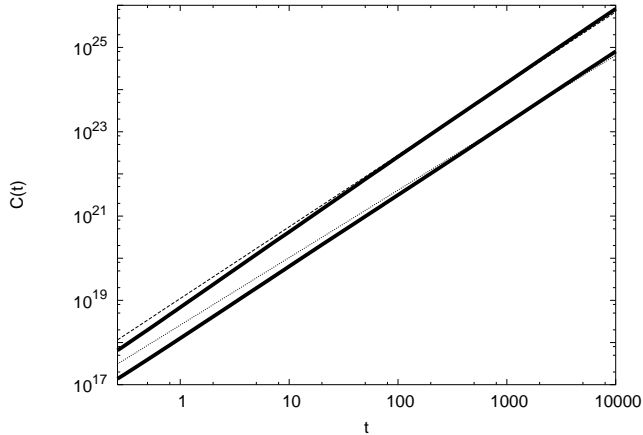,width=6.0cm,angle=-90}\label{fig1}
\caption{Log-log plot of $C(t)$ versus $t$ for values of 
$\alpha = 0.6$ (above) and $0.8$ (below). The thin dotted straight lines 
are fits to the data (in thick lines) with a slope equal to $2H = 2-\alpha/2$ 
as given by Eq.  \ref{hurst2}, and at large times $t$ the data approach 
the straight lines. The data curves are scaled by suitable factors for visual
clarity.  }
\end{center}
\end{figure}

\begin{figure}
\begin{center}
\psfig{file=\DIR/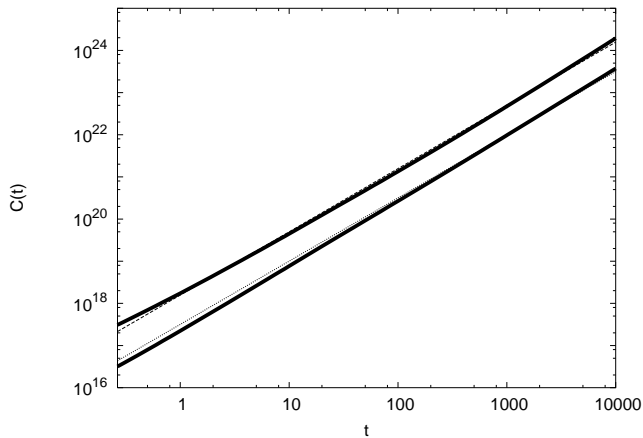,width=6.0cm,angle=-90}
\caption{Log-log plot of $C(t)$ versus $t$ for values of $\alpha = 3.5$ 
(above) and $1.5$ (below).  The thin dotted straight lines are fits to the 
data (in thick lines) with a common slope equal to $2H = 3/2$, as in Eq. 
\ref{hurst2}. Both the curves approach this slope at large times. The 
curves are scaled for visual clarity.} 
\label{fig2}
\end{center}
\end{figure}

We note that to see the correct asymptotic behaviour, one has to
explore distance scales $y \ll a$.  We consider a grid along the
$y$-direction with grid spacing $\Delta y$, extending from $-N =
-y_{max}/\Delta y = -500000$ to $N = y_{max}/\Delta y = 500000$.  At
each point of this grid we choose $v(y)$ by a discrete Fourier
transform of the random numbers ${\tilde v}(k_m)$ obtained for all
$k_m$ (with $k_m = -N, \dots, N$).  Once a set of $\{v(y)\}$ is thus
chosen, they remain fixed at all times during different thermal
histories. This set $\{v(y)\}$ constitutes a particular realization of
disorder. Finally one performs the disorder average ${\overline
{\left(\dots \right)}}$ by averaging over various realizations of the
set $\{v(y)\}$. In all our simulations we have averaged over $100$
realizations of disorder using $1000$ thermal realizations for each
set of disorder.

There are further, three important things to be noted. Firstly, for
$\alpha < d$, the role of cutoff $a$ is unimportant and hence it may
be safely ignored. We have verified for $d=1$ that the results are
unchanged in comparison to finite $a$, by taking $a \rt 0$, which is
equivalent to setting $c(ka)=1$ in Eq. \ref{vv_k}. However, the role
of $a$ in the $\alpha>d$ regime cannot be overemphasized. The cutoff
$a$ has to be chosen such that $a > \Delta y$. We have used
$a=10~\Delta y$ in our simulations.  For low values of the ratio
$a/{\Delta y}$ , the expected behaviour of $C(t)\sim t^{2-d/2}$ is not
seen. The second point is that the value $k=0$ leads to numerical
difficulties for $\alpha<d$. The latter was taken care of by using a
very small finite value of $k$ in place of $k=0$.  Finally we note
that the two integrals in the constant $c(ka)$ are rapidly decaying
integrals and better results are obtained by using very small grids of
integration.

In figure 2, we see that for different $\alpha < d (=1)$, the
incremental correlation function $C(t) \sim t^{2H}$ with exponent $H =
1 - \alpha/4$, as predicted by Eq. \ref{hurst2}.  In figure 3, we see
that for different $\alpha > d (=1)$, again $C(t) \sim t^{2H}$, but now
with $2H = 2 - d/2 = 3/2$.

\begin{figure}
\begin{center}
\psfig{file=\DIR/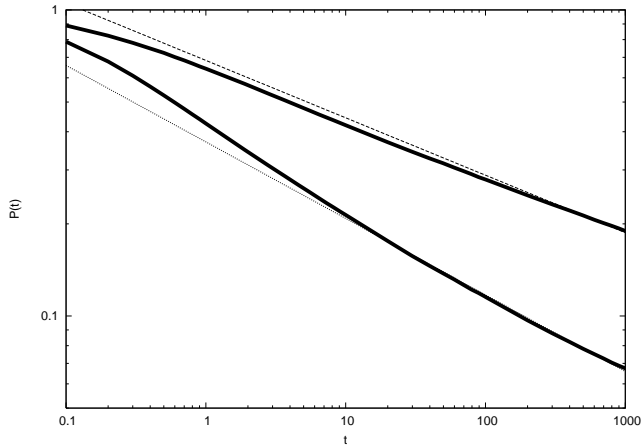,width=6.0cm,angle=-90}
\caption{Log-log plot of $P(t)$ versus $t$ for the values of $\alpha = 0.75$ 
(above) and $3.5$ (below). The straight lines are fits to the data with a 
slope equal to $-\theta$ where $\theta = 1-H$ (see Eq. \ref{theta}).}
\label{fig3}
\end{center}
\end{figure}

We directly measured the averaged persistence probability $P(t)$ in
our simulations. The procedure is described clearly in
\cite{Krug1}. Our results are shown in figure 3.  We have shown $P(t)$
for two different $\alpha$ values: for $\alpha < d (= 1)$, the power
law exponent $\theta = \alpha/4$ as expected, while for $\alpha > d
(=1)$ we have $\theta = d/4 = 1/4$.

\section{Remarks and Conclusion}
\label{discuss}

In this paper, we have derived the incremental two-time correlation
function $C(t)$ and persistence probability $P(t)$ of the longitudinal
$x$-coordinate of a random walker subjected to a quenched random {\it
power law} correlated velocity field. This problem is a variant of the
original MdM model, now with long range velocity correlation.  The
main result is shown in figure \ref{fig4}. There are three distinct
regions in the parameter space of $d$ versus $\alpha$. $(i)$ For low
dimensions ($d < 2$) and large power law exponents ($\alpha > d$), the
behaviour of $C(t)$ and $P(t)$ are same as that of the short-range
correlated MdM model \cite{sn1}. Thus this regime is {\it short-range
disorder} dominated. $(ii)$ On the other hand for strong long-range
disorder i.e. $\alpha < {\rm min}[d,2]$, the temporal behaviour of
$C(t)$ and $P(t)$ depend on $\alpha$ continuously, so that long-range
correlation really has a new effect. The latter regime is {\it
long-range disorder} dominated.  $(iii)$ Finally for $d \geq 2$ and
$\alpha \geq 2$, the Hurst exponent is $1/2$ , i.e., the same as that
of a simple random walker (SRW).  Quenched disorder is irrelevant, and
this regime is thus {\it thermal noise} dominated. Further, at the
boundaries in the Fig. \ref{fig4} denoted by the dotted lines, there
are logarithmic corrections to the power law expressions for $C(t)$
and $P(t)$.

We note that the the quenched random velocity field is having an
effect on the temporal behaviour of a SRW, only when the disorder is
sufficiently strong and transverse dimension $d$ is small. For
short-range correlated MdM model \cite{sn1} the $H$ changes only when
the number of transverse directions $d < 2$. It is as if the velocity
kicks in the $x$-direction fail to have much impact when the particle
has many more transverse directions to explore. In this paper we find
that for $\alpha < {\rm min}[d,2]$, i.e. for stronger correlated or
longer ranged disorder, the short-range correlated MdM model exponents
give way to new exponents. Thus even if transverse dimensions are large,
sufficiently long-range disorder makes the random walker 
move super-diffusively. For high dimension ($d > 2$) and weak 
disorder ($\alpha > 2$), the quenched noise looses out to the 
thermal noise, and the asymptotic temporal properties are that of 
simple Brownian motion. 

We also note that the short-distance cutoff scale $a$ of the power
law, plays a crucial role in this problem.  For the case $\alpha > d$,
numerically, one has to be very careful to explore distance scales
much smaller compared to the power-law cutoff scale $a$. It is one of the
most important aspects of this regime that very short distance
correlations seem to control the long time behaviour. For the regime
of $\alpha < d$, since correlations at scale $y^{\prime} > a$ determine the 
asymptotic behaviour rather than $y^{\prime} < a$, the cutoff $a$ may 
be ignored.  
 
\section{Acknowlegements}

We are thankful to S.N. Majumdar for reading the manuscript carefully
and giving the very important suggestion of analyzing the case of
$\alpha > d$ with a finite cutoff. It led to the correct understanding
of the exponents in that regime. We are also thankful to A. Sain for
some helpful suggestions.  D. Das acknowledges the Seed grant
no. $03ir051$ of I.I.T. Bombay, India, for financial support.

\end{document}